\documentclass[iop]{emulateapj}

\slugcomment{Version: \today}
\usepackage{times}
\usepackage{hyperref}
\def\F{{\it Fermi}-LAT }

\def\NS{{\it NuSTAR }}

\shorttitle{Inverse Compton X-ray emission from Mrk~421}  
\begin{document}

\title{Inverse Compton X-ray Emissions from TeV blazar Mrk~421 \\
during a Historical Low-Flux State Observed with {\it NuSTAR}}

\author{
Jun Kataoka\altaffilmark{1} and \L ukasz Stawarz\altaffilmark{2}}
\altaffiltext{1}{Research Institute for Science and Engineering, Waseda University, 3-4-1, Okubo, Shinjuku, Tokyo 169-8555, Japan}
\altaffiltext{2}{Astronomical Observatory of Jagiellonian University, ul. Orla 171, 30-244 Krakow, Poland}

\email{email: {\tt kataoka.jun@waseda.jp}}

\begin{abstract}
We report on the detection of excess hard X-ray emission from the TeV BL Lac object 
Mrk~421 during the historical low-flux state of the source in January 2013. 
\NS observations were conducted four times between MJD~56294 and MJD~56312 with 
a total exposure of 80.9\,ksec. The source flux in the $3-40$\,keV range was 
nearly constant except for MJD~56307, when the average flux level increased by 
a factor of three. Throughout the exposure, the X-ray spectra of Mrk~421 were 
well represented by a steep power-law model with a photon index of 
$\Gamma \simeq 3.1$, although a significant excess was noted above 20\,keV 
in the MJD~56302 data when the source was in its faintest state. 
Moreover, Mrk~421 was detected at more than the $4 \sigma$ level in the 
$40-79$\,keV count maps for both MJD~56307 and MJD~56302 but not during the 
remaining two observations. The detected excess hard X-ray emissions connect 
smoothly with the extrapolation of the high-energy $\gamma$-ray continuum of 
the blazar constrained by \F during the source quiescence. These findings 
indicate that, while the overall X-ray spectrum of Mrk~421 is dominated by the 
highest-energy tail of the synchrotron continuum, the variable excess hard 
X-ray emission above 20\,keV (on the timescale of a week) is related 
to the inverse Compton emission component. We discuss the resulting constraints 
on the variability and spectral properties of the low-energy 
segment of the electron energy distribution in the source. 
\end{abstract}


\section{Introduction}

Blazars are a subclass of radio-loud active galactic nuclei (AGN) for which 
non-thermal jet emissions are relativistically beamed along the line of sight
 \citep[for a review see, e.g.,][]{Begelman84,Urry95}. Thus, pronounced variability 
on timescales of as short as days and even hours is often observed in 
various energy bands \citep[e.g.,][]{Wagner95,Ulrich97}. 
The broadband electromagnetic spectra of blazars consist of two characteristic bumps in the $\nu - \nu F_{\nu}$ representation: one extending from radio to optical/X-ray frequencies
 and the other one peaking around high-energy $\gamma$-ray photon energies. The measured 
polarization in the radio and optical bands indicates that the low-energy 
spectral component is due to the synchrotron emissions of ultrarelativistic 
electrons; the $\gamma$-ray continuum is instead 
most widely believed to result from inverse Comptonization (IC) of 
various soft photon fields. The source of the seed photons for the IC process 
can be either the synchrotron emission of the jet itself \citep[``Synchrotron self-Compton'' model, or SSC for short; e.g.,][]{Jones74,Marscher80,Band85}, or the thermal 
emissions of circumnuclear gas and dust \citep[e.g.,][]{Dermer93,Sikora94,Inoue96}. 
There is an ongoing debate on the exact localization of the dominant blazar emission 
zone, with current estimates ranging from hundreds of gravitational radii from 
central supermassive black holes up to parsec-scale distances, as well as on the dominant particle acceleration processes involved, with different scenarios including mildly 
relativistic internal shocks \citep[e.g.,][]{Boettcher10,Mimica12,Saito15}, 
magnetic turbulence \citep[e.g.,][]{Yan13,Asano14,Zheng14,Kakuwa15}, 
or even relativistic magnetic reconnection \citep[e.g.,][]{Narayan12,Biteau12,Sironi15}.

Mrk~421 (R.A.~=~11h04m27.3s, DEC~=~+38d12m32; $z$~=~0.031) is a nearby and bright 
blazar classified as a typical ``high-frequency-peaked'' BL Lac (HBL) object 
whose synchrotron and IC emission components extend up to X-ray and 
very high energy (VHE) $\gamma$-ray photon energies, respectively. 
It is in fact the first extragalactic source detected in the TeV range \citep{Punch92}. 
As such, Mrk~421 is one of the most comprehensively studied HBLs through a number 
of observation campaigns. In particular, recent multi-wavelength campaigns, 
including \F, have provided the first ever complete coverage of the $\gamma$-ray 
continuum of the source in its low-activity/quiescence state \citep{Abdo11b}. 
Interestingly, the fractional variability amplitude $F_{\rm var}$, when plotted 
as a function of frequency, also reveals a double-peak structure echoing 
the spectral energy distribution (SED) of Mrk~421: $F_{\rm var}$ rises significantly 
from the radio towards the X-ray frequencies, decreases over the \F band, and 
finally rises again in the TeV regime \citep{Abdo11b,Balokovic16}. 
This finding seems to imply that, in the framework of the ``homogeneous one-zone'' 
emission model, the high-energy tail of the electron energy distribution 
(synchrotron X-rays and TeV $\gamma$-rays emitted via the IC process) is 
highly variable, while the low-energy segment of the electron energy distribution 
is relatively steady. The alternative interpretation is a stratified (multi-zone) 
blazar emission model with highly variable high-energy emissions produced in 
distinct (more compact) emission sites compared with lower-energy emissions, which 
vary more slowly.

\begin{figure*}[t!]
\begin{center}
\includegraphics[angle=0,scale=0.8]{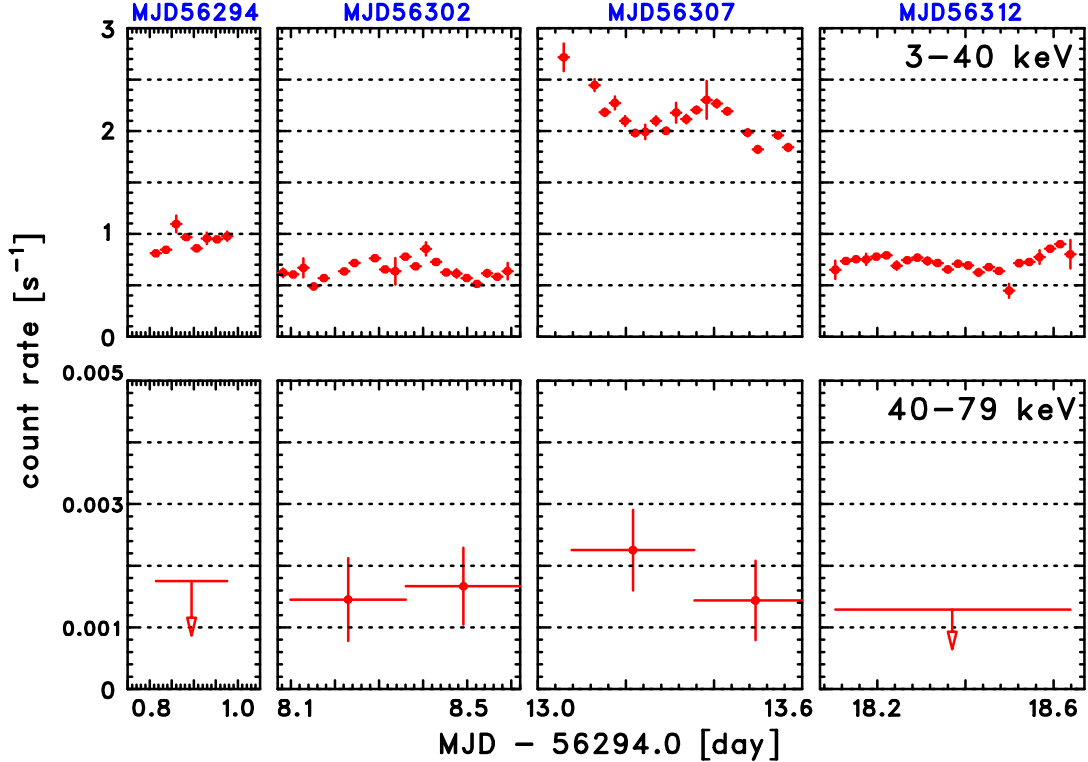}
\caption{Overall hard X-ray variability of Mrk~421 observed with \NS during 
the historical low-flux state of the source between MJD~56294 (January 12, 2013) 
and MJD~56312 (January 20, 2013). Count rates in the $3-40$\,keV range 
(top panels) and $40-79$\,keV range (bottom panels) are summed from 
both FPMA and FPMB (after background subtraction) and binned in 2\,ks and 20\,ks 
intervals, respectively. For a detection significance of less than 
$2 \sigma$, upper limits are given for the $95 \%$ confidence level. }
\label{fig:lightcurve}
\end{center}
\end{figure*}

In this context, broadband X-ray observations of Mrk~421 may in principle 
provide unique clues on the variability and spectral properties of both the 
low- and the highest-energy segments of the electron energy distribution 
because the synchrotron continuum of the blazar falls exponentially 
around photon energies of several/tens of keV (at least during the low-activity states).
This reflects the high-energy cutoff in the underlying electron energy 
distribution, so the radiative output of the source in the hard X-ray range
 may be dominated by the low-energy tail of the IC emission component. 
However, in the case of BL Lacs in general and HBLs in particular, including the 
X-ray brightest source Mrk~421, X-ray data above 10\,keV are deficient and 
available typically only for isolated flaring periods 
\citep[see][]{Ushio09,Ushio10,Abdo11b}. This is because previous hard X-ray 
observations relied on ``non-imaging'' detectors for which the sensitivity 
was insufficient to detect the source in its low-activity states. 
The Nuclear Spectroscopic Telescope Array mission \citep[\NS;][]{Harrison13} 
is the first focusing high-energy X-ray telescope in orbit and operates in the 
band from 3 to 79\,keV. Thanks to its imaging capability, \NS 
probes the hard X-ray sky with a more than 100-fold improvement in sensitivity. 

\citet{Paliya15} and \citet{Sinha15} presented the \NS observations of Mrk~421 
during flaring epochs. In this paper, we revisit the archival Mrk~421 data 
provided by \NS in January 2013 and discussed previously by \citet{Balokovic16} 
when the source was in a historical low-activity state corresponding 
to an extrapolated $2-10$\,keV energy flux 
of $\simeq 4 \times 10^{-11}$\,erg\,cm$^{-2}$\,s$^{-1}$.

\begin{figure*}[t!]
\begin{center}
\includegraphics[angle=0,scale=0.31]{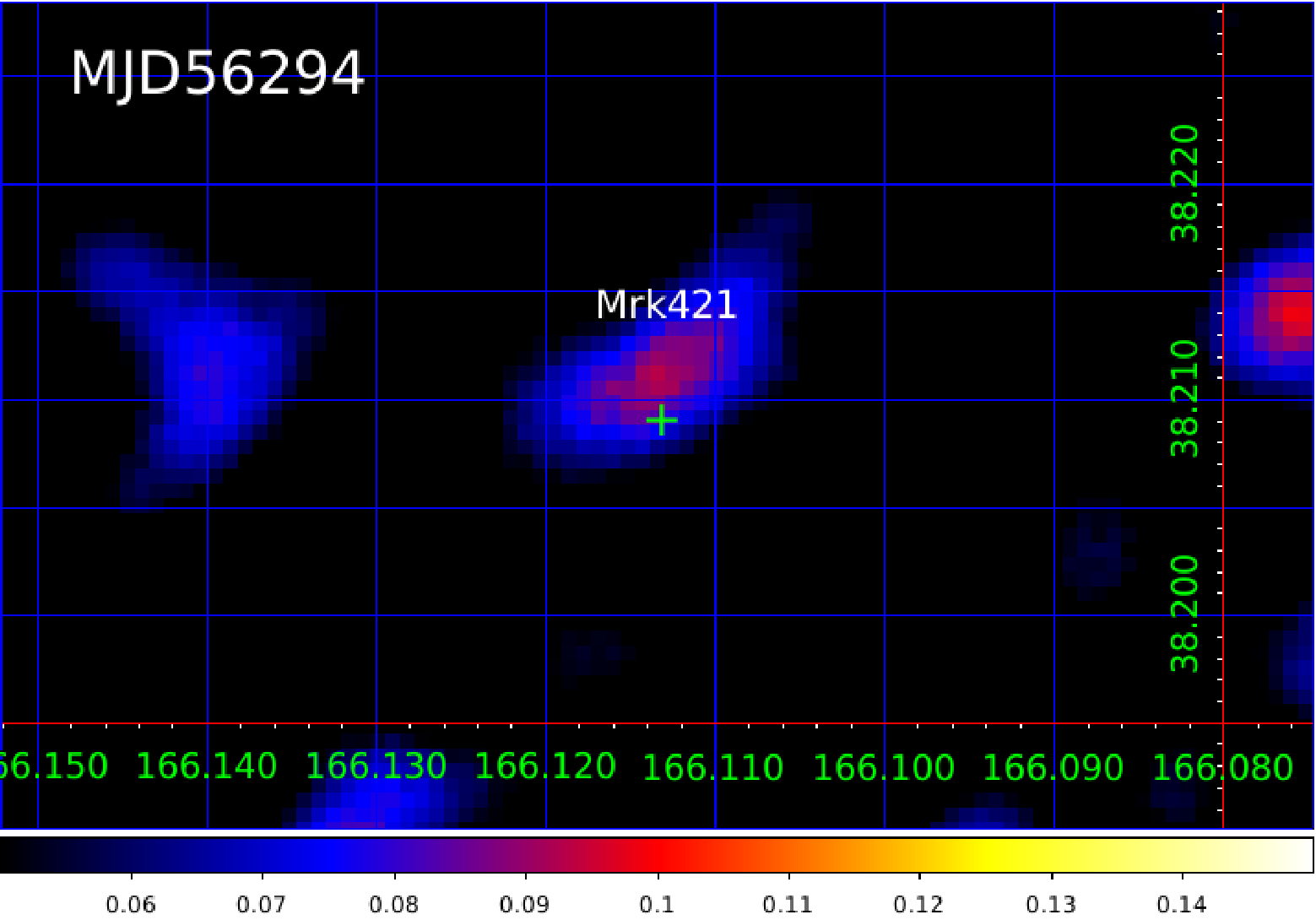}
\includegraphics[angle=0,scale=0.31]{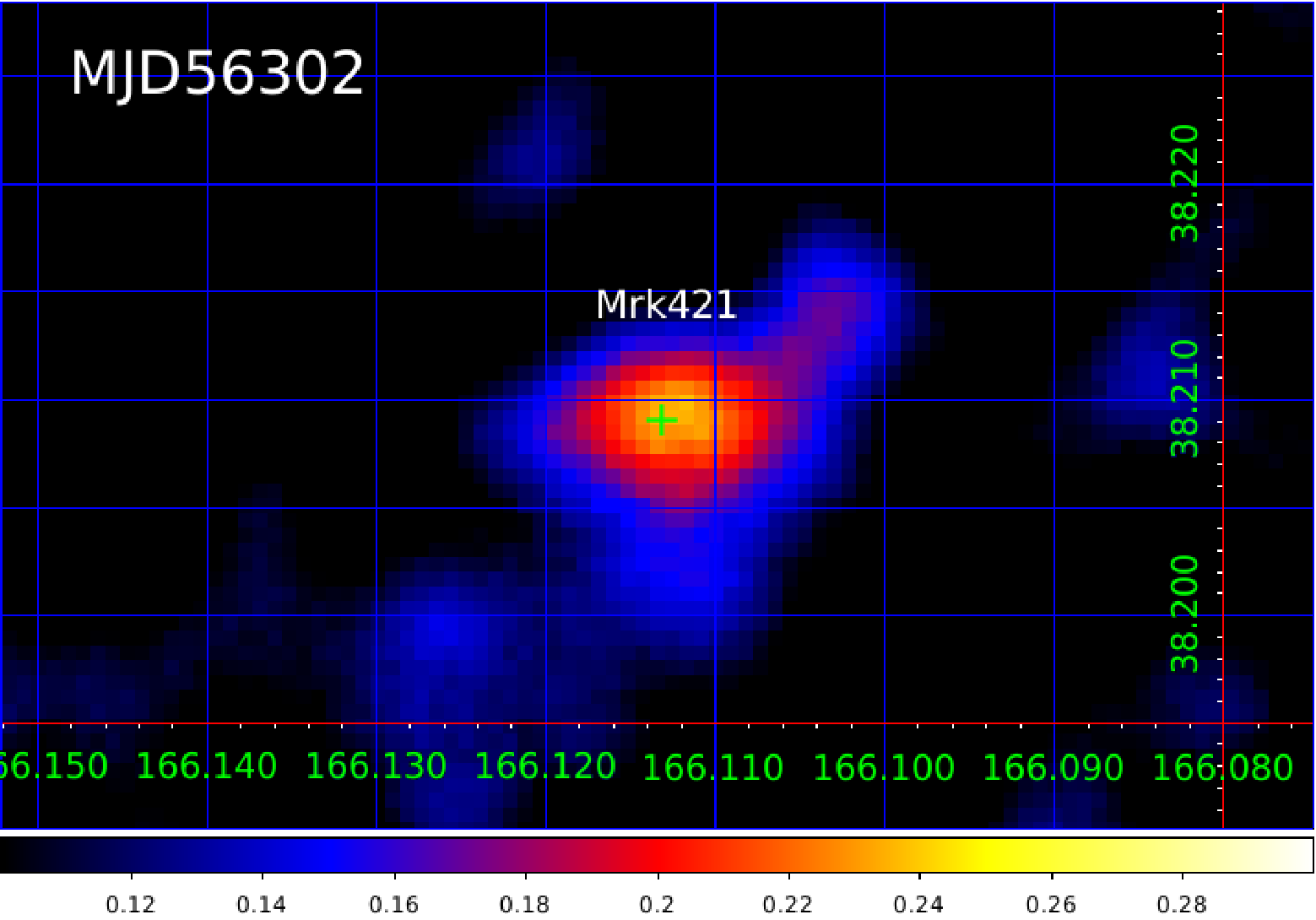}
\includegraphics[angle=0,scale=0.31]{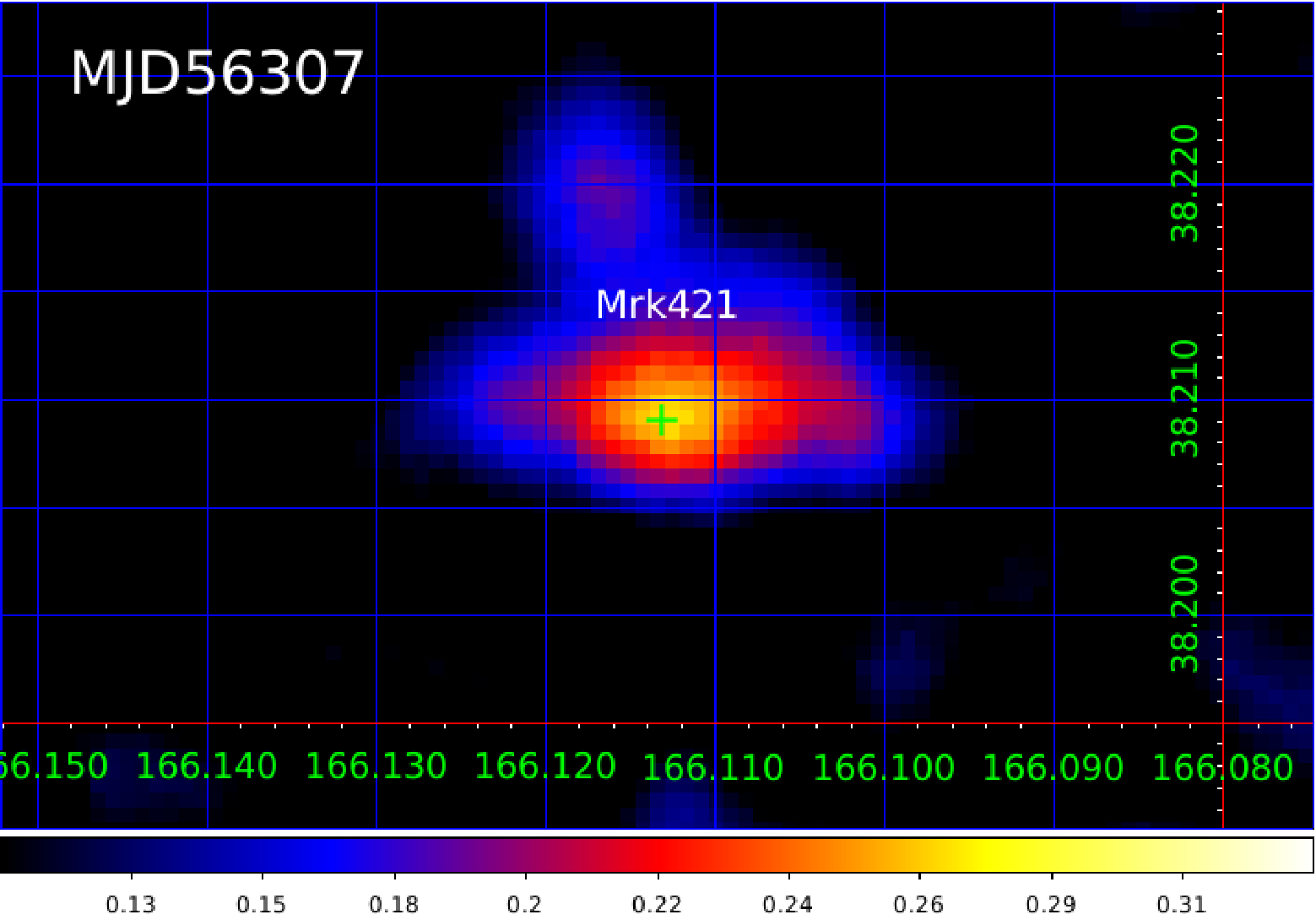}
\includegraphics[angle=0,scale=0.31]{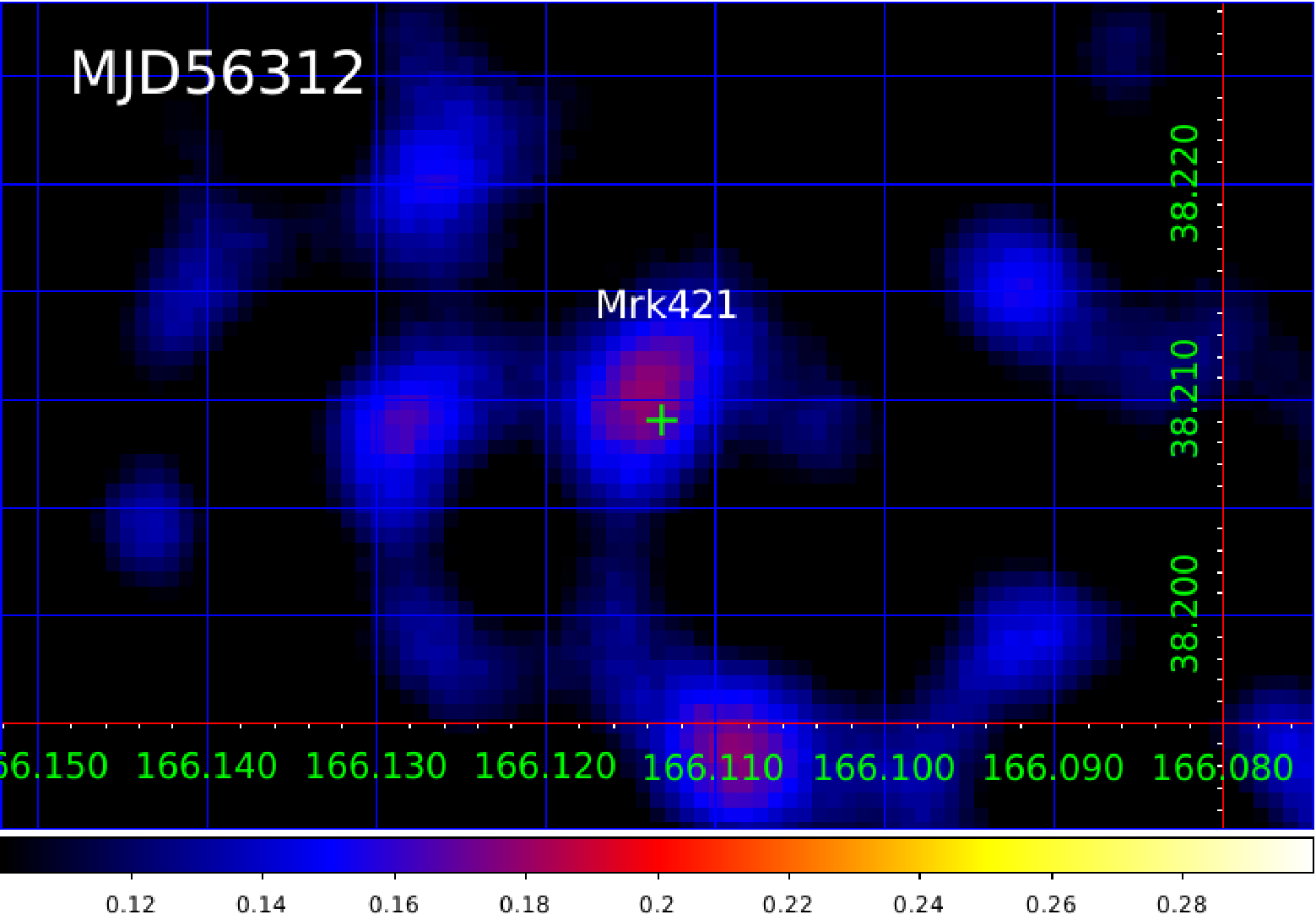}
\caption{Temporal variations in the \NS images of Mrk~421 in the 
$40-79$\,keV energy range. The images denotes the relative excess of photon counts 
(arbitrary units indicated in the color bar corrected for difference in 
exposures; see Table\,\ref{table1}) smoothed with a Gaussian kernel of $19.6''$. 
Photon counts from FPMA and FPMB are summed. Note the source was significantly detected 
at more than the $4 \sigma$ level only in MJD56302 and MJD56307, which consistent 
with the source light curve given in Figure\,\ref{fig:lightcurve}.}
\label{fig:images}
\end{center}
\end{figure*}

\section{Observations and Results} 
\subsection{Observation and Data Reduction}

Mrk~421 was selected as a representative of the HBL class of blazars 
and thus has been observed frequently with \NS since January 2013, partly in the 
framework of extensive multi-frequency campaigns. The total exposure 
time over 88 orbits amounts to more than 250\,ks, including two calibrating 
observations conducted in July 2012 \citep{Balokovic16}. A typical \NS observation 
spans 10 hrs; hence, the actual exposure for each observation, after only ``good 
time intervals'' are selected to eliminate the passage of the South Atlantic Anomaly 
and orbital modulation of visibility, ranges from 10\,ks to 25\,ks. All observations 
were conducted with two co-aligned independent telescopes called Focal Plane 
Module A and B (FPMA and FPMB). Throughout the observations, variations in the 
count rate of  more than two orders of magnitude were observed. 
It has been argued that the synchrotron emissions of the blazar account 
for all the flux detected up to 79\,keV during the flaring states 
\citep{Paliya15,Sinha15}. Here, we focus on the historical low-flux state of Mrk~421, 
which was witnessed by \NS in January 2013 
(see Table\,\ref{table1} for the logs of the analyzed observations).

The archival \NS data were downloaded from NASA's HEASARC 
interface\footnote{\tt http://heasarc.gsfc.nasa.gov/docs/archive.html} and reduced by 
using the \texttt{NuSTARDAS} software, which is included as part of the \texttt{HEAsoft} 
package version 6.17. For the temporal and spectral studies presented in the following 
sections, the X-ray events were extracted from a circular region with a radius of $30''$ 
and centred on the source position, whereas the background was accumulated in an 
annulus with inner and outer radii of $30''$ and $70''$, respectively.  
We carefully checked that other choices of  source and background regions 
did not affect the analysis results presented in the next sections within uncertainties 
of $1 \sigma$. The response files were generated by using the standard \texttt{nupipeline}
 and \texttt{nuproducts} scripts. Version 20151008 of the CALDB files were used in 
this study. Even in the historical low-flux state of the target, the extracted 
X-ray count rates were well above the background level up to 40\,keV and 
almost equal to the background level above 40\,keV. With  good 
characterization of the \NS background, which left the uncertainty at a 
$1 \%$ level \citep{Wik14}, the gathered Mrk~421 data could therefore 
be used for spectral modeling up to the high-energy end of the \NS band at 79\,keV.

\subsection{Analysis Results}

Figure\,\ref{fig:lightcurve} shows the count rate variations during the \NS 
observations between January~2 (MJD~56294) and January~20 (MJD~56312). 
The X-ray counts from both FPMA and FPMB detectors are summed, and the background is 
subtracted. The light curves are shown separately for the two different 
energy bands: $3-40$\,keV (top panels) with 2\,ks binning and $40-79$\,keV 
(bottom panels) with 20\,ks binning. The light curves indicate that the  
$3-40$\,keV source flux was nearly constant except for MJD~56307 when a 
pronounced variability was observed and the average flux increased by a factor 
of three. In addition, note that the source was detected at more than the 
$3 \sigma$ level in the 40$-$79 keV band  
not only in MJD~56307 but also in MJD~56302, when an 
average $3-40$\,keV count rate was lowest. For the remaining 
two observations, we provide the corresponding $95 \%$ confidence level upper limits. 

Figure\,\ref{fig:images} compares the temporal variations in the \NS 
X-ray images of Mrk~421 in the $40-79$\,keV energy range. These were reconstructed 
by using the sum of the FPMA and FPMB data in MJD~56294, 56302, 56307 and 56312, 
respectively. The images were smoothed by a Gaussian kernel of $19.6''$ 
and color-coded; the given color bar indicates a  relative excess of photon counts 
in arbitrary units for which the maximum value is corrected for different exposure 
times of each observation (see Table\,\ref{table1}). Note that the target was clearly 
detected at more than the $4 \sigma$ level only in MJD~56302 and 56307, which is 
consistent with the source light curve given in Figure\,\ref{fig:lightcurve}. 
A slightly higher significance for the images ($> 4\sigma$) compared to the light curves 
($> 3\sigma$) is because source counts were extracted from a circular region with a 
radius of $30''$ for Figure\,\ref{fig:lightcurve}, whereas all of the detected 
photons were used for Figure\,\ref{fig:images} against the point spread function 
of \NS. A positive detection of the source above 40\,keV during the extremely 
low-flux state in MJD~56302 but not during similarly low-flux states 
in MJD~56294 and 56312 indicates that the excess hard X-ray emission is 
variable on the time scale of a week.

\begin{deluxetable*}{lccccccc}[t!]
\tabletypesize{\scriptsize}
%\rotate
\tablecaption{Summary of \NS observations and analysis of Mrk~421 }
\tablewidth{0pt}
\tablehead{
\colhead{Start Time} &  \colhead{Exposure} & model &\colhead{$\Gamma_1$$^d$} & \colhead{$E_{\rm brk}$$^e$} & \colhead{$\Gamma_2$$^f$} & \colhead{$3-79$\,keV energy flux} & \colhead{$\chi^2$/dof}\\
\colhead{(MJD)} &  \colhead{(ksec)} & \colhead{} & \colhead{}  & \colhead{(keV)}  & \colhead{} & \colhead{(10$^{-12}$\,erg\,cm$^{-2}$\,s$^{-1}$)} & \colhead{}
}
\startdata
56294.78 & 9.2 & PL$^a$ & 3.10$\pm$0.03 & $-$ & $-$ & 44.4$\pm$0.6 & 170.8/161\\
         &  & BKN-PL$^b$ & 3.25$\pm$0.05 & 7.23$\pm$0.66 & 2.87$^{+0.06}_{-0.07}$ & 46.4$\pm$4.2 & 155.1/159\\
         &  & PL+PL$^c$ & 3.32$\pm$0.07 & $-$ & 1.7$^f$ & 48.4$\pm$0.9 & 158.1/160\\
\tableline
56302.05 & 22.6 & PL & 3.06$\pm$0.02 & $-$ & $-$ & 31.0$\pm$0.3 & 262.3/249\\
         &  & BKN-PL & 3.08$\pm$0.02 & 17.8$^{+2.1}_{-1.6}$ & 2.08$^{+0.26}_{-0.28}$ & 34.0$\pm$1.3 & 246.3/247\\
         &  & PL+PL & 3.11$^{+0.04}_{-0.03}$ & $-$ & 0.17$^{+0.71}_{-0.79}$ & 35.3$\pm$0.5 & 247.9/247\\
\tableline
56307.04 & 24.2 & PL & 3.03$\pm$0.01 & $-$ & $-$ & 107.6$\pm$0.5 & 467.4/456\\
         &  & BKN-PL & 2.96$\pm$0.02 & 7.83$^{+0.6}_{-0.5}$ & 3.19$^{+0.04}_{-0.03}$ & 105.1$\pm$3.3 & 429.6/454\\
\tableline
56312.10 & 25.0 & PL & 3.07$\pm$0.02 & $-$ & $-$ & 36.4$\pm$0.3 & 295.3/290\\
         &  & BKN-PL & 3.07$\pm$0.02 & 28.8$^{+9.2}_{-5.2}$ & 1.78$^{+0.81}_{-1.78}$ & 38.2$\pm$1.2 & 292.5/288\\
\enddata 
\tablecomments{
$^a$: The power-law model, \textsc{wabs*pegp}, where the absorption column density was fixed to the Galactic value in the direction of the source, $N_{\rm H} = 1.5 \times 10^{20}$\,cm$^{-2}$.\\
$^b$: The broken power-law model, \textsc{wabs*bkenpower}, where the absorption column density was fixed to the Galactic value.\\
$^c$: The double power-law model, \textsc{wabs*(pegp+pegp)}, where the absorption column density was fixed to the Galactic value.\\
$^d$: The power-law index for the PL model, the low-energy power-law index for the BKN-PL model, or finally the power-law index for the first PL components in the PL+PL model.\\
$^e$: The break energy for the BKN-PL model in keV.\\
$^f$: The high-energy power-law index for the BKN-PL model, or the power-law index for the second PL components for the PL+PL model.\\
}
\label{table1}
\end{deluxetable*}

For the spectral fitting, we binned the source light curve to a minimum of 40 counts per 
bin to enable the $\chi^2$ minimization statistics. Table\,\ref{table1} lists the 
resulting fitting parameters for all of the analyzed observations, for which 
errors are quoted at the $1\sigma$ confidence level. The Galactic absorption column 
density toward Mrk~421 was taken to be $N_{\rm H} = 1.5 \times 10^{20}$\,cm$^{-2}$  
\citep{Elvis89}. Figure\,\ref{fig:spectra} shows the FPMA (black symbols) 
and FPMB (red symbols) spectra during MJD~56302 plotted against the best-fit 
power-law (PL), broken power-law (BKN-PL) in which high-energy photon index is harder than that in the low-energy part, and double power-law (PL+PL) models 
within the energy range of $3-79$\,keV. 
The residuals in the figure corresponding to the PL fit 
(best-fit photon index $\Gamma = 3.10\pm0.03$ with $\chi^2$/dof of 262.3/249) 
indicate that the source spectrum exhibits significant hard excess emissions 
above 20\,keV. The fit was significantly improved in the cases of both the 
BKN-PL and PL+PL models ($\chi^2$/dof of 246.3/247 and 247.9/247 dof, respectively). 
Thus, the improvement in the $\chi^2$ statistic as measured with the $F$ static value of 
8.0 was significant at more than the $99.9 \%$ level using the $F$-test. 
The spectral hardening was statistically less significant 
in the MJD~56294 and 56312 data (see Table\,\ref{table1}).

\section{Discussion and Conclusion}

\begin{figure}[t!]
\begin{center}
\includegraphics[angle=0,width=\columnwidth]{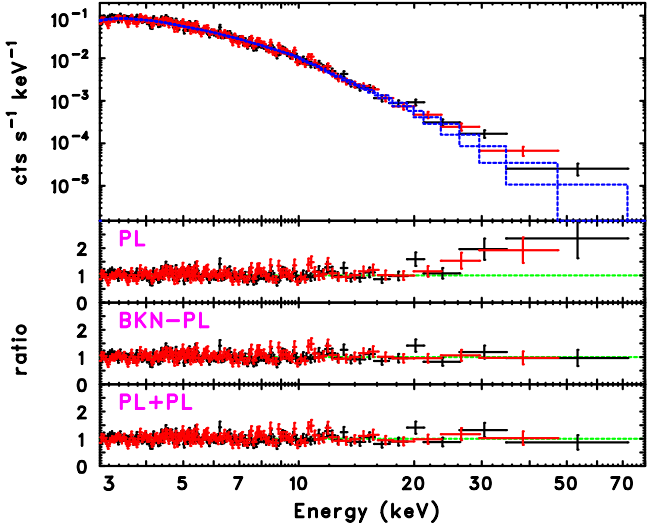}
\caption{\NS spectrum of Mrk~421 from the MJD~56302 data. 
The upper panel presents the FPMA (black) and FPMB (red) data plotted 
against an absorbed power-law model with the photon 
index $\Gamma \simeq 3.1$ and the Galactic 
column density of $N_{\rm H} = 1.5 \times 10^{20}$\,cm$^{-2}$, 
fitted over the $3-79$\,keV band. The bottom panels show the 
data/model ratio residuals for the power-law fit (PL), 
broken power-law fit (BKN-PL), and double power-law fit (PL+PL). 
Deviations above 20 keV are clearly seen in the PL fitting, as detailed in the text. } 
\label{fig:spectra}
\end{center}
\end{figure}

In the previous sections we presented an analysis of the archival \NS observations 
of Mrk~421 in January 2013 when the overall X-ray flux was particularly low. 
In MJD~56032, the $2-10$\,keV energy flux estimated from the extrapolation 
of the PL fit (photon index $\Gamma \simeq 3.1$) was 
$(4.02\pm0.05)\times 10^{-11}$\,erg\,cm$^{-2}$\,s$^{-1}$, which is the lowest 
ever reported in the literature for the source. The detection of Mrk~421 
above 40~keV in such a low flux state manifested in a concave broadband X-ray 
spectrum of the target with excess hard X-ray emissions dominant at photon energies 
above 20\,keV  (Table\,\ref{table1} and Figure\,\ref{fig:spectra}). However, this 
excess was not found in the MJD~56294 or 56312 data when the source was in a 
similarly low-flux state (Figures\,\ref{fig:lightcurve} and \ref{fig:images}), 
which implies a variability of the hard X-ray continuum on the timescale of a week. 

\begin{figure}[t!]
\begin{center}
\includegraphics[angle=0,width=\columnwidth]{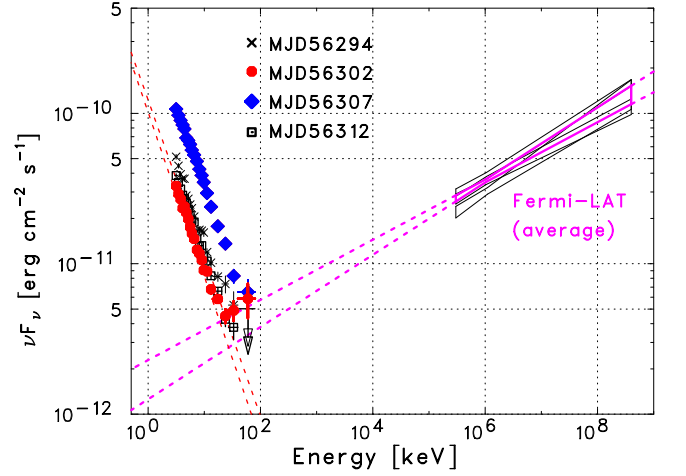}
\caption{Broadband X-ray (\NS; this paper) to high-energy 
$\gamma$-ray \citep[\F;][]{Abdo11b} spectra of Mrk~421. 
Bow-ties plotted in black denote the \F data taken in different 
multi-frequency campaigns; the average source spectrum during the 
period of a source quiescence (from 2008 August 5 to 2010 February 20) is shown as 
a magenta bow-tie. Dashed magenta lines mark the extrapolation of the 
best-fit power-law model to the average \F data ($\gamma$-ray photon index
 of $1.78 \pm 0.02$). }
\label{fig:SED}
\end{center}
\end{figure}

To investigate the physical origin of the detected excess emission, we compiled 
the broad-band X-ray (\NS; this work) and the high-energy $\gamma$-ray 
\citep[\F;][]{Abdo11b} data in a $\nu - \nu F_{\nu}$ representation, 
as shown in Figure\,\ref{fig:SED}. 
Bow-ties plotted in black denote the \F data taken during different multi-frequency 
campaigns; the average source spectrum during the period of a source quiescence 
(from 2008 August 5 to 2010 February 20) is shown as a magenta bow-tie. 
The dashed magenta lines mark the extrapolation of the best-fit power-law model 
applied to the average \F data \citep[$\gamma$-ray photon index of $\Gamma_{\gamma} \sim 1.78 \pm 0.02$;][]{Abdo11b}, with $1 \sigma$ uncertainty. 
The excess hard X-ray emissions detected in the analyzed \NS data agreed well with
 the extrapolation of the average \F spectrum. This indicates that the quiescence 
continuum of Mrk~421, which extends from photon energies of tens of keV up to tens of GeV,
 corresponds to a single IC emission component and is well-described by a 
power-law model with a photon index of $\simeq 1.8$. 

Interestingly, similar spectral upturns related to the synchrotron/IC crossover 
have been detected below 10\,keV in several other blazars classified as 
``low-frequency-peaked'' BL Lacs \citep[LBLs; see, e.g.,][]{Tagliaferri00,Tanihata03,Wierzcholska16}
 but never in HBLs either below or above 10\,keV. Further novelty here is that the 
high-energy excess feature found in the \NS data for Mrk~421 is variable on the 
 time scale of a week; no evidence for such variability in the low-energy 
segment of the IC emission continuum has been reported for other BL Lacs 
in the literature. 

An alternative explanation for the observed hard X-ray excess of the target is a 
spectral pile-up in the electron distribution $N'_e(\gamma_e)$, where 
$\gamma_e$ is the electron Lorentz factor, forming temporarily at the highest 
energies.\footnote{Here we do not consider a possibility that the observed hard 
X-ray excess is related to the bulk Comptonization of the accretion disk emission 
by cold electrons within the innermost parts of Mrk~421 jet, since the corresponding 
bulk-Compton spectral features have been predicted for (and possibly even observed in) 
luminous blazars of the ``flat-spectrum-radio-quasar'' type \citep[see, e.g.,][and references therein]{Sikora97,Kataoka08}, and not for BL Lac objects characterized 
by low accretion rates and hence radiatively inefficient accretion disks.} 
This feature may appear due to either (i) a continuous (stochastic) acceleration 
of electrons limited by radiative losses \citep[e.g.,][]{Stawarz08}, or 
(ii) the reduction of the IC cross section in the Klein-Nishina 
regime \citep[e.g.,][]{Moderski05}. In the case of (i), the pile-up bump 
appears at the maximum electron energies for which the acceleration timescale 
equals the radiative loss timescale at the limit for the perfect confinement 
of electrons within the emission zone (i.e., no particle escape). However, the power-law 
tail at lower electron energies has to be relatively flat in such a scenario, 
$s \equiv - d \ln N'_e(\gamma_e)/d \ln \gamma_e = 0-1$, which disagrees with the 
observation in Mrk~421. Similarly, in the case of (ii), the necessary condition for 
the formation of a pronounced spectral hardening in the electron energy distribution 
is the dominance of IC cooling over the synchrotron one and a relatively narrow 
(within the frequency range) seed photon distribution. 
These requirements contradict the conditions expected for the Mrk~421 jet in 
its quiescence (for which the synchrotron losses dominate the IC ones 
and the soft photon distribution for the IC scattering is provided by the broadband 
synchrotron emission of the jet itself). Therefore, while also considering 
the extrapolation of the \F spectrum of the source to the X-ray regime, 
we argue that the hard X-ray excess found in the \NS data indeed represents 
the low-energy tail of the IC (SSC) component.

From the observed SSC photon energy of $h \nu_{\rm SSC} \simeq 20$\,keV 
and a steep broadband power-law electron energy distribution, 
the corresponding minimum electron Lorentz factor is roughly
\begin{equation}
\gamma_{e,\rm min} \sim 10^3 \, \left(\frac{B'}{0.1{\rm G}}\right)^{-1/4} \left(\frac{\delta}{10}\right)^{-1/4} \, .
\end{equation}
The radiative cooling timescale for such low-energy electrons dominated by 
the synchrotron process is then
\begin{equation}
\tau_{\rm cool}(\gamma_{e,\rm min}) \sim 10^7 \, \left(\frac{B'}{0.1{\rm G}}\right)^{-7/4} \left(\frac{\delta}{10}\right)^{-3/4}\,\rm{s} \, ,
\end{equation}
which is much longer than a week for the comoving magnetic field 
intensity $B' \lesssim 0.1$\,G emerging from the one-zone SSC model 
applied to the quiescence SED of Mrk~421 \citep{Abdo11b}. Hence, the variability 
of the low-energy electrons implied by the \NS observations analyzed in this 
paper have to be related to dynamical changes within the blazar emission zone 
for which the shortest timescale is given by the light crossing time $R/c$. 
Interestingly, this would agree with the emission region size assumed 
in the SED model of \citet{Abdo11b}. In particular, one has
\begin{equation}
R \simeq 0.06 \, \left(\frac{t_{\rm var}}{1 {\rm week}}\right) \, \left(\frac{\delta}{10}\right)\, \rm{pc} \, , 
\end{equation}
which implies the distance of the emission zone from the active nucleus 
$r \sim \Gamma_j \, R \sim 0.6$\,pc for the anticipated conical jet geometry 
with the opening angle $\sim 1/\Gamma_j$ and the jet bulk Lorentz 
factor $\Gamma_j \sim \delta \sim 10$. 
This is also in accord with the detailed analysis of the overall 
variability of Mrk~421 at X-ray frequencies, which implies that the power in 
the intraday flickering of the source is small \citep{Kataoka01,Isobe15}.

In addition, the results of our \NS data analysis revealed that the electron 
energy distribution in Mrk~421 during the source quiescence is well-represented by 
a relatively steep power law with the energy index of 
$s \simeq 2 \Gamma_{\gamma} -1 \sim 2.6$, which extends from electron energies 
of at least $\gamma_e \sim 10^3$ up to $\gamma_e \sim 10^6$ (in the jet rest frame). 
This is an important finding because in such a case the bulk of the jet kinetic 
energy is carried by those low-energy electrons 
(assuming no significant proton content). The exact value of the electron spectral index 
is also important based on the most recent results for the kinetic simulations of 
relativistic magnetic reconnection process. In other words, the power-law energy 
spectra formed within the reconnection sites are steep ($s > 2$ only if the 
jet magnetization is low \citep{Sironi14,Guo15}. Again, low magnetization of 
the emission region would be in accord with the model parameters emerging from a simple 
one-zone SSC models \citep{Abdo11a,Abdo11b}. However, note that this conclusion is 
with regard to only the slowly variable/quiescence emission component because the production of rapid high-amplitude flares in blazar sources may proceed under very different conditions, and may involve very different electron spectra.

\acknowledgments 

\L .S. was supported by Polish NSC grant DEC-2012/04/A/ST9/00083.

{}

\end{document}